# The Impact of COVID-19 on Urban Energy Consumption of the Commercial Tourism City

Dongdong Zhang[a#], Hongyi Li[b#], Hongyu Zhu[a#], Hongcai Zhang[b*], Jonathan Goh [a], Hui Liu[a], Man Chung Wong[b]

a School of Electrical Engineering, Guangxi University, Nanning, China
b State Key Laboratory of Internet of Things for Smart City, University of Macau, Macau, China
*Corresponding author: hczhang@um.edu.mo
[#]Dongdong Zhang, Hongyi Li and Hongyu Zhu contributed equally to this paper.

**Highlights:**

- Analysed the trend of energy production and consumption in a typical commercial tourist city represented by Macao during the COVID-19 pandemic.
- Completed energy consumption data in various fields in Macao during the COVID-19 pandemic.
- Compared the energy consumption features of different commercial tourist cities in the world during the COVID-19 pandemic.

**Abstract:**

In 2020, the COVID-19 pandemic spreads all over the world. In order to alleviate the spread of the epidemic, various blockade policies have been implemented in many areas. This has led to sluggish demand in the world's major economies, a sharp drop in trade index, and negative growth in energy consumption. Comparing with large comprehensive cities or countries, the unique economy structure of Macao makes it a typical case of analysing the energy consumption of the commercial tourism cities during the epidemic period. In order to formulate a better epidemic prevention policy for urban energy consumption of the commercial tourism cities, this paper first summarised the major statistics of energy supply and demands in Macao before and during the epidemic period based on actual data. Then, the characteristics of the energy consumption in different sectors of Macao, including hotel, transportation, tourism culture and public utilities, are analysed in detail. Finally, the energy consumption features of commercial tourism cities represented by Macao are compared with other typical countries/regions (e.g. Italy, the United States, Japan and Brazil). These analyses demonstrate the impact of the COVID-19 on energy consumption in commercial tourism cities, which provide insights for the government or energy providers to formulate their policies to adapt to this pandemic.

Keywords: The COVID-19 pandemic; Energy management; Energy structure characteristic; Urban Energy Consumption; Macao

# 1. Introduction

The outbreak of the COVID-19 pandemic in 2020 has widely affected more than 200 countries and regions. By the end of 2020, more than 1.8 million people died [1]. COVID-19 has strong infectivity and long incubation period, resulting in the exponential growth and waveform expansion of the epidemic. To contain the spread of the virus, many countries and regions have adopted unprecedented quarantine measures, including closing cities, suspending work, forbidding gathering, working and studying from home etc., which changed the way of human life and production [2]. From the perspective of energy, transportation has a major impact on energy [3-4]. The strict border management and restrictions on going abroad implemented by countries all over the world have made air transportation suffer from significant economic depression [5]. It is predicted that the passenger volume will reduce by 861 million to 1292 million person-times in 2020 [6], which will seriously affect the energy consumption characteristics of cities. With the rapid emergence of virtual commerce, education and social networking platforms, the COVID-19 pandemic has also brought about significant digital transformation. Various uncertainties brought about by the epidemic situation have stimulated further exploration of future technologies, and significantly increased the combination and innovation of various industries and AI technologies in the future [7].

Interestingly, due to the reduction of most energy demand and global economic activities, the global environment has been evidently improved in the short term [8]. The reduction of population mobility and industrial production has greatly reduced air pollution and greenhouse gas emissions [9], resulting in the highest reduction of carbon emissions in the past decade [10]. The COVID-19 pandemic has a tremendous impact on the short-term social economic development and people's life, further limiting the consumption demand and the export trade dominated by manufacturing industry [11-12]. Such impacts have resulted in the reduction of electricity demand and brought unprecedented challenges to the power industry that plays a supporting role in the development of human society.

To retard the spread of the epidemic, city closure policy was implemented, which changed the structure of energy demand during the epidemic period. On the one hand, as most factories and enterprises are shut down, the decline of industrial and commercial demand have made the overall energy demand of all parts of the world present a significant downward trend [13-14]. On the other hand, due to the implementation of policies such as restrictions of going out, many regions began to implement measures such as work-from-home and remote study, thereby the residential energy consumption increased significantly [15]. The change of energy consumption structure features reshaped the load curves correspondingly. Affected by the epidemic and travelling restriction policy, people's lifestyles and social modes have changed greatly. For example, the proportions of remote working and information sharing in human life are continuously increasing. The way of working from home makes people's time distribution of energy use more flexible, which causes a time swift in the peak-valley curve [16]. As the peak period of energy consumption moves back as a whole, the load peak-valley difference between working days and rest days also decreases.

With the change of energy demand structure, the energy structure in supply side also changed during the epidemic period. Affected by the COVID-19 pandemic, the overall demand for electricity has declined, resulting in the reduction of thermal power generation and the uptrend of multiple distributed renewable energy generation in many regions [17-18]. These trends also reflect the urgent need of exploring various cutting-edge technologies of renewable energy, smart grid and efficient energy storage [19]. Due to the strong intermittence and volatility of the renewable energy [20], the increase of the renewable energy utilization during the epidemic period will inevitably pose more challenges to power system operation [21]. In addition, the changing characteristics of energy demands also lead to the increase of the randomness and uncertainty, which make load forecasting more challenging than that in the past. As a result, economic and secure planning and operation of multi-energy systems now face more challenges [22].

At the same time, affected by the COVID-19 pandemic, public utilities and enterprises face with two major problems. 1) Income shrinkage and cash loss caused by reduced demand and falling prices [23]. 2) Increasing cost of maintaining the normal operation of the power system under such special circumstances [24]. Moreover, due to the problems of reduced demand and supply chain termination, many state-investigated energy projects and plans for new facilities and infrastructure are delayed or shelved, such as India's renewable energy project [25] and the US capacity expansion plan [26]. Meanwhile, various bans have played a positive role in reducing urban carbon emissions and improving the environment. However, it is unsustainable and impractical to mitigate climate change by reducing human activities. A more effective way is the implementation of the clean transformation of energy type [27]. Therefore, it is essential to have a deeper understanding of the changing trend of power energy industry during the epidemic period and analyse the new trend of energy development in the future, which can response to climate change requirements, sustainable development and energy transformation needs of various countries.

For commercial tourism cities, whose energy sources usually rely on the purchase from adjacent areas, the COVID-19 outbreak' impacts have higher complexity due to the uniqueness of the economic structure. A better understanding of the energy consumption of the commercial tourism cities such as Macao under the epidemic is of great significance and can provide a reference for policy makers to formulate better policies. Based on the evolution trend, prevention and control efforts, and the resumption of work, this paper analyses the energy consumption of Macao during the epidemic period. First, it gives a macroscopic analysis of the overall situation of the impact of COVID-19 on Macao's energy, and then analyses the overall situation of the new energy consumption characteristics, and conducts a detailed analysis of the impact of the COVID-19 pandemic in different industries of Macao (such as hotels, cultural events, economic events, etc.). Then, the detailed comparison of Macao's energy characteristics with four large comprehensive countries' energy characteristics (Italy, the United States, Japan and Brazil) is performed, which shows the unique characteristics of Macao's energy consumption during the epidemic period. Based on the results of the analysis of energy utilization characteristics, it provides a better reference for the subsequent development of better epidemic control measures and recovery work in commercial tourism cities like Macau.

## 2. Macro Analysis of COVID-19's Influence on Macao's Energy Consumption

The epidemic prevention and control strategies, energy consumption adjustments and fluctuations in various industries have brought challenges to Macao's energy industry. The demand and supply structure of power and natural gas, power generation and discharge of power plants, and integrated energy services are directly or indirectly affected. According to Macao government's database, the main performance and changing trend of its energy industry were analyzed.

### 2.1. Electricity Demand and Supply

As a result of the COVID-19 pandemic, the closure of most sectors of industry, commerce and tourism has reduced the total power consumption in Macao by 7% year-on-year. Fig. 1 shows the power consumption trend in Macao. As a special administrative region of China, Macao's power supply mainly comes from Mainland China [28]. In addition, the local power generation includes waste incineration centre, natural gas, fuel oil and photovoltaic power generation. The decline in overall energy consumption during the epidemic period has led to a 7% drop in electricity purchases from Mainland China. However, due to the rise in power generation using natural gas, the electricity production of local power plants increased by 3%. In addition, Macao's total electricity consumption in the first quarter of 2020 decreased by 20% quarter-on-quarter, with 95.6% of the electricity purchased from Mainland China. Electricity generated by local power plants dropped by 95%, including 62.7% from natural gas and 37.3% from fuel oil. Meanwhile, waste incineration centres and photovoltaic power generation dropped by 24% and 25%, respectively.

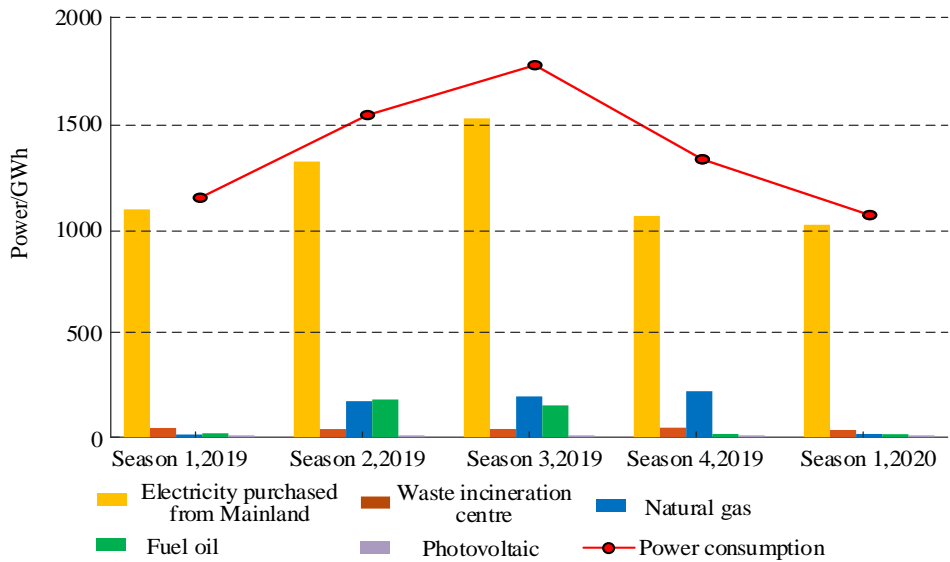

Fig. 1 Power consumption trend in Macao

*2.2. Natural Gas Demand and Supply*

The overall energy consumption characteristics of natural gas is shown in Fig. 2. In the first quarter of 2020, Macao imported 7.43 million m$^3$ of natural gas, decreased by 88% quarter-on-quarter while increased by 13% year-on-year. In the same quarter, 7.45 million m$^3$ of natural gas was consumed, of which 46.7% was used for power generation and 53.3% was used for urban gas supply. Macao' unique geographical environment has made the tourism service industry, which is the lifeline of the city's economy, vigorously developed [29]. During the epidemic period, the restriction of tourism caused a huge loss of passenger flow in Macao, which significantly reduced the energy consumption of tourism service. In contrast, home office and remote teaching policy have increased the energy demand in residential areas. Fig. 3 shows the consumption characteristics of natural gas used for urban gas supply. In the first quarter of 2020, the gas consumption of Macao's urban pipeline network decreased by 25% quarter-on-quarter, of which the sales of business and services account for 89.1%. Residential gas supplied by gas facility operators accounted for 9.8% and public welfare gas accounted for 1.2%. Compared to the same quarter last year, the overall consumption dropped by 25%. The data of the natural gas consumption of several commercial, tourism-related customers was collected, including one restaurant and six hotels, to analyze the impact of COVID-19 epidemic on the tourism industry. The 99$^{th}$ percentiles of natural gas load consumption of each customer, from Jan 2018 to July 2020, are calculated, which filter out the abnormal peaks caused by drift and can be regarded as the corresponding customer's maximum simultaneous natural gas load. Then, the *coincidence factors (CFs)* of each customer defined by the ratio of each customer's maximum simultaneous natural gas load to the total installed natural gas load capacity are also calculated, as shown in Table. I. The kernel density estimations of the data in year 2018, 2019 and 2020, which visualizes the distribution of the natural gas load consumption of the two types of customers (restaurant and hotel), and the CFs of each customers are also compared in Fig. 4.

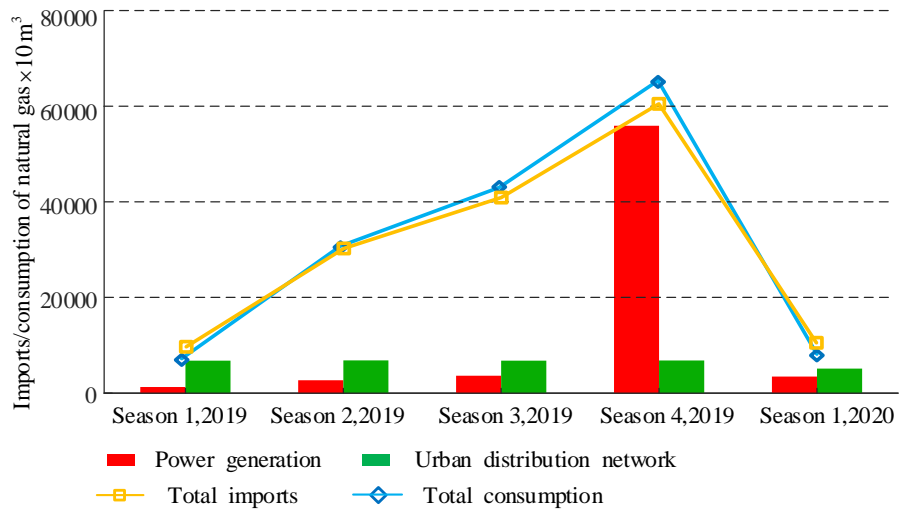

Fig .2 Overall energy consumption characteristics of natural gas

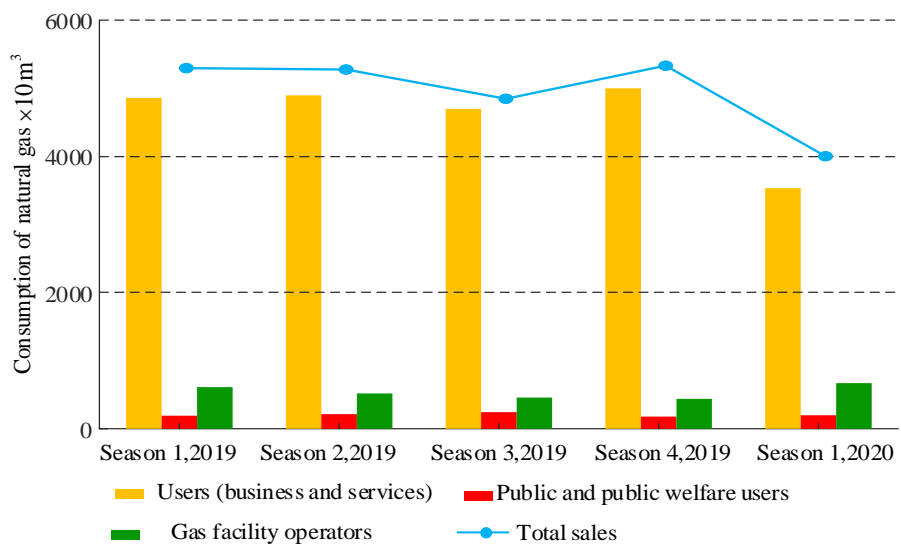

Fig. 3 The consumption characteristic of natural gas used for urban gas supply

As implied by the 99$^{th}$ percentiles and CFs in Table I, in most of the using scenarios, the actual natural gas load is usually far below the installed natural gas load capacity for each customer. The CF of the restaurant is below 0.5 and the CFs of the hotels are all below 0.24 in 2018 to 2020. As shown in Fig. 4(a) and Fig. 4(b), the gas load levels and CFs of the restaurant and the hotels in 2020 is obviously lower than the ones in 2018 and 2019, resulting in a higher peak at the low load value area. This reduction implies that during the epidemic period, the business of the tourism is bleak. Since the gas load have reduced under the COVID-19 epidemic, the capacity of the current distribution system has the potential to serve more need of commercial customers in the future. Instead of upgrading the system in terms of supply capacity, the gas suppliers could invest more on the maintenance and intellectualization of the system.

Table I  The installed capacity, 99th percentiles and CFs of the commercial gas loads

| Customer | Installed Capacity (m³/h) | 99th percentile of Actual Gas Load (m³/h) | | | Coincidence factors | | |
| --- | --- | --- | --- | --- | --- | --- | --- |
| | | 2018 | 2019 | 2020 | 2018 | 2019 | 2020 |
| Restaurant | 41.2 | 18.2259 | 19.0115 | 14.2039 | 0.4424 | 0.4614 | 0.3448 |
| Hotel 1 | 2211.1 | 401.6863 | 343.9133 | 307.0790 | 0.1817 | 0.1555 | 0.1389 |
| Hotel 2 | 2572.5 | 321.1200 | 303.7600 | 248.7255 | 0.1248 | 0.1181 | 0.0967 |
| Hotel 3 | 3219.7 | 802.1347 | 752.7692 | 653.4925 | 0.2491 | 0.2338 | 0.2030 |
| Hotel 4 | 760.9 | 103.8752 | 73.9823 | 72.9926 | 0.1365 | 0.0972 | 0.0959 |
| Hotel 5 | 4473.2 | 943.5000 | 655.5136 | 525.4876 | 0.2109 | 0.1465 | 0.1175 |
| Hotel 6 | 2590.7 | 531.2900 | 303.0277 | 179.7056 | 0.2051 | 0.1170 | 0.0694 |

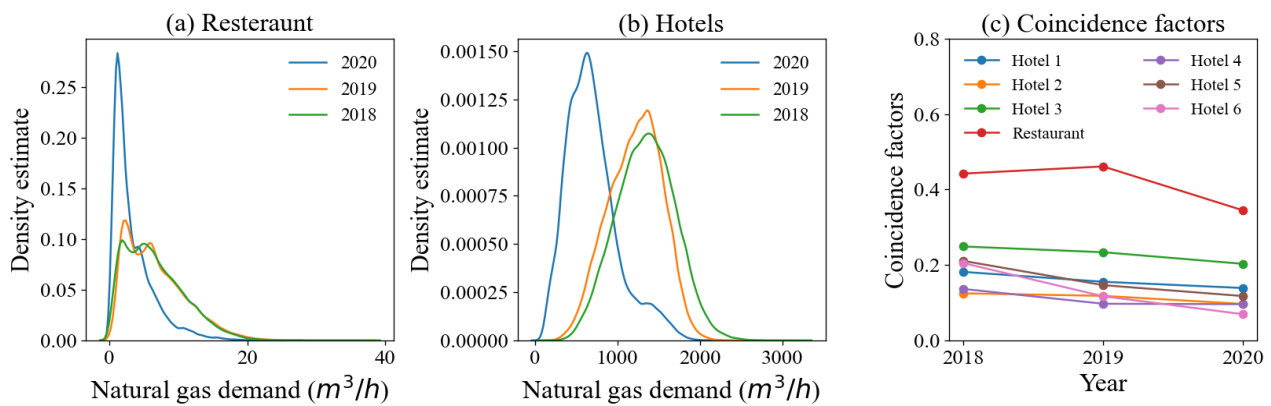

Fig. 4  The kernel density estimation and CFs of the commercial loads

*2.3. Power Generation and Discharge of Power Plant*

During the epidemic period, the unique energy consumption characteristics and fluctuating loads have caused a certain degree of change in Macao's overall energy supply structure. The shutdown of most factories and traffic restrictions greatly reduced the emission of various pollutants. The first outstanding performance was the significant decrease of greenhouse gas emissions, which is shown in Fig. 5. In the first quarter of 2020, the carbon dioxide ($CO_2$) and sulphur dioxide ($SO_2$) emissions from power plants in Macao were reduced by 75% and 43%, respectively. The discharge of biochemical oxygen demand, chemical oxygen demand and suspended solids in wastewater was evidently reduced. Although the COVID-19 pandemic caused the reduction of urban emissions and effectively improved the environment, this was achieved on the basis of restricting human economic activities and hindering social development, which is unsustainable. In the future, what is really conducive to environmental development is to accelerate the transformation of energy structure and technological innovation.

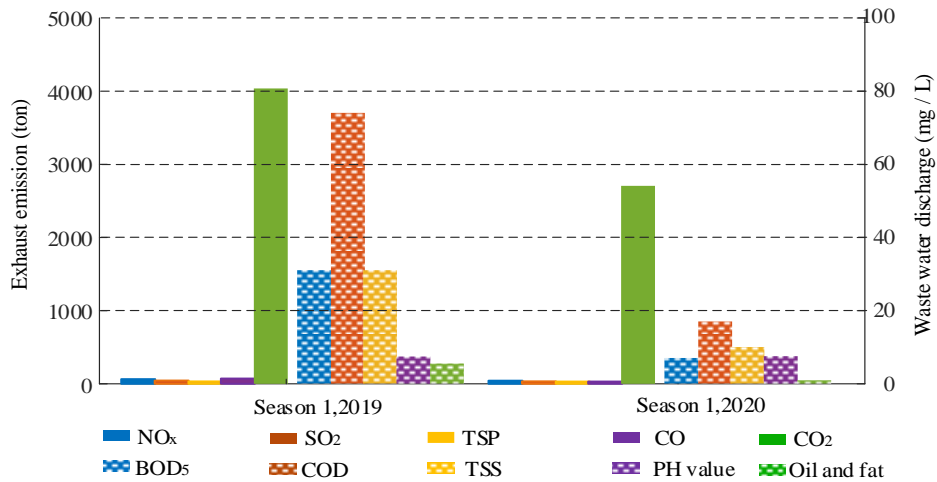

Fig. 5  Discharge of power plant waste

## 2.4. Average Cost of Electricity

Macao's average electricity costs in the first and last seasons of 2019 and first season of 2020 as shown in Fig. 6. As can be seen from the figure, the cost of electricity purchased in Macao in the first quarter of 2020 has dropped by 4% compared to the first and forth quarters of 2019. The cost of power generation from oil and natural gas increased by 46% compared to the fourth quarter in 2019. In addition, the cost of power generation from oil and natural gas in Macao increased by 35% and 28% year-on-year respectively. Although the cost of electricity generation from oil and natural gas has increased by a large proportion, most electricity in Macao comes from the purchase of Mainland. Therefore, the overall average cost of electricity in Macao in the first quarter of 2020 reduced by 2% quarter-on-quarter and 3% year-on-year.

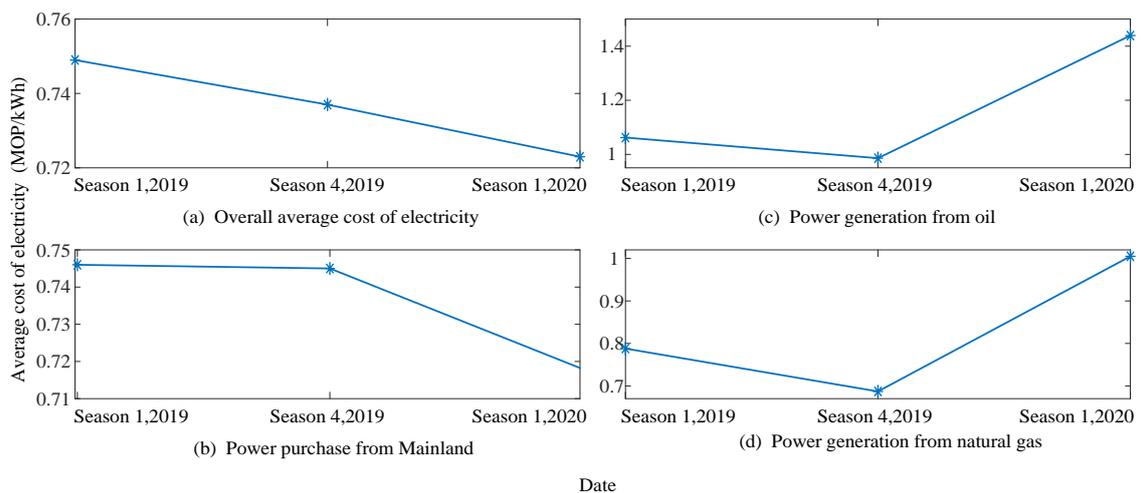

Fig. 6  Macao's electricity costs and investment

## 2.5. Power Supply Customer Service

In the short term, the sluggish tourism industry during the epidemic has seriously affected Macao's overall energy consumption. Some office workers carried out the strategy of working from home, which has led to an increase in residential electricity consumption, while industrial, commercial and government agencies' energy consumption dramatically droped. Fig. 7 reflects the overall electricity sales trend in Macao. By the end of March, there were 268322 electricity customers in Macao, of which residential and commercial customers accounted for 86.6% and 11.0%, respectively. The overall electricity sales in the first quarter of 2020 decreased by 18% compared with the previous quarter, of which commercial users decreased significantly by 22%, residential users decreased by 2%, and government agencies and industry users decreased by 21% and 18%, respectively. In addition, comparing with the same quarter last year, the overall electricity sales fell by 7%, resulting from the decrease of consumption of all users except residential users.

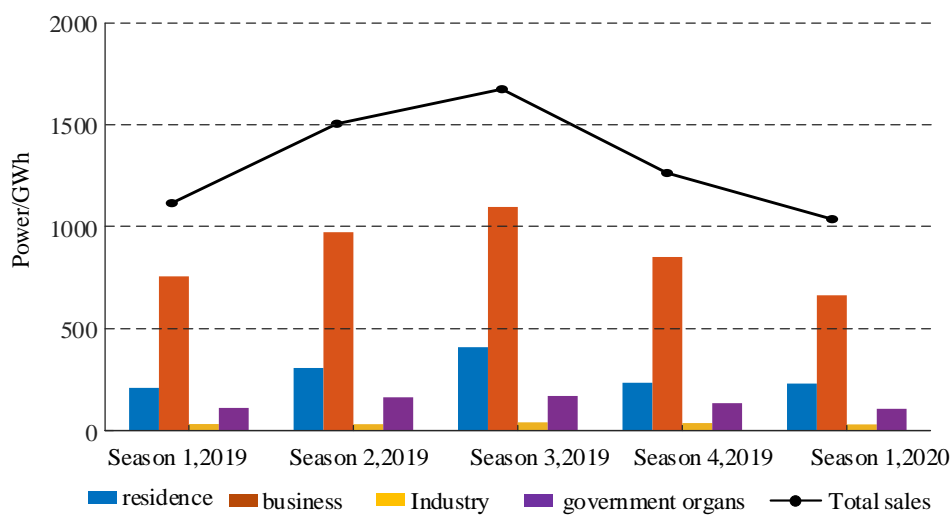

Fig. 7 The overall electricity sales trend in Macao

## 3. Impact of COVID-19 on Energy Consumption in Different Fields of Macao

From the end of January 2020, the government in Macao started to pose restriction on commercial events, especially lottery industry. On January 27$^{th}$, people who have travelled to Hubei province in the last 14 days are confined to enter casinos, which is recognized as the start of lockdown in policy aspect. The government furthery closed most of the indoor entertainment venues on February 4$^{th}$, including all the casinos, cinemas, game-centres et al. This strict lockdown measure caused an obvious drop in the electricity demand of relative industries. Such restriction measures were gradually removed in late February and March, due to the government's quick response and control to the COVID-19 pandemic. The industries began to reopen from then on. But in April the reviving of economy is interrupted because of the second outbreak of the epidemic caused by the imported cases. This section adopts the anonymous smart meter data collected by the electricity utility company, i.e., Companhia de Electricidade de Macau (CEM), to investigate the detailed implication of COVID-19 for electricity demand in different industries. It is worth noting that only part of the electricity consumed were recorded by the smart meters. Thus, in this paper, we mainly focus on the difference of electricity consumption between 2019 and 2020 and the overall trend observed. The discussion about the numerical amount is not included.

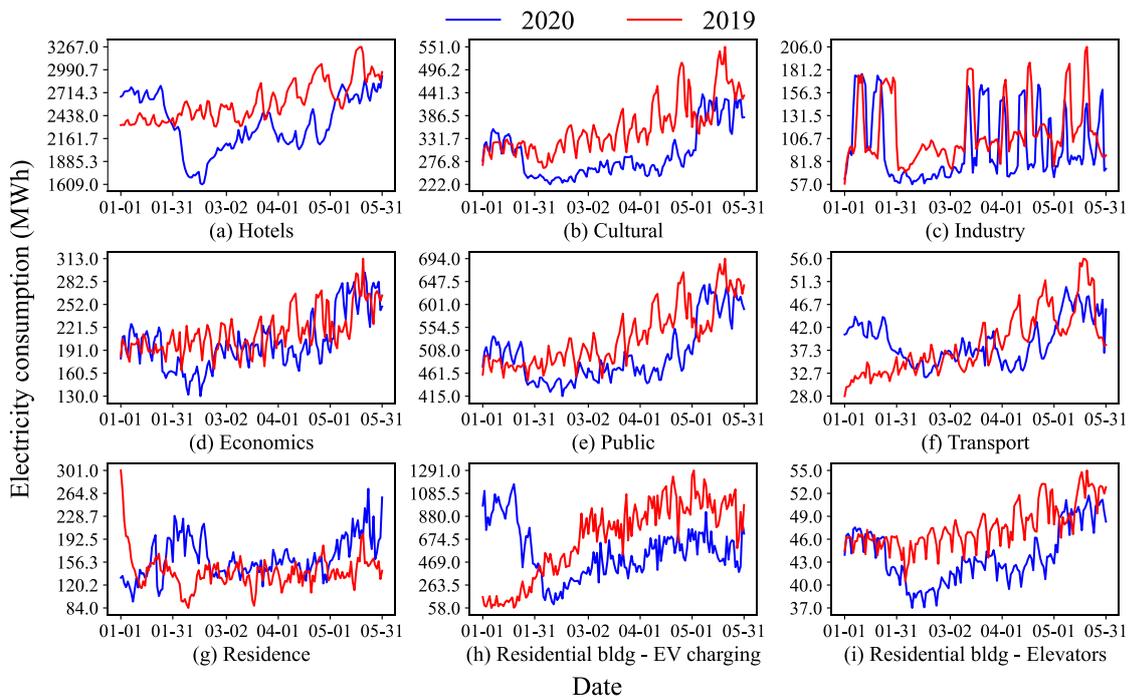

Fig. 8 Daily electricity consumption of different sectors during the first 5 months of 2019 and 2020

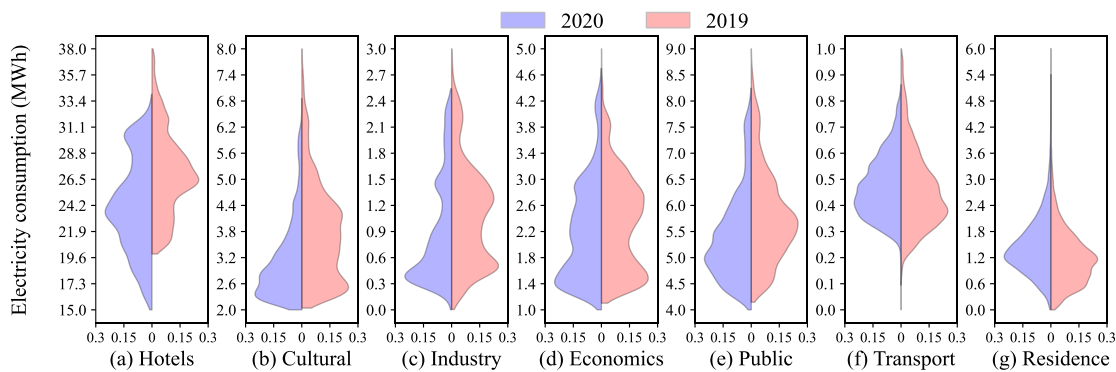

Fig. 9 Distribution of electricity consumption (per 15 minutes) of different sectors during the first 5 months of 2019 and 2020

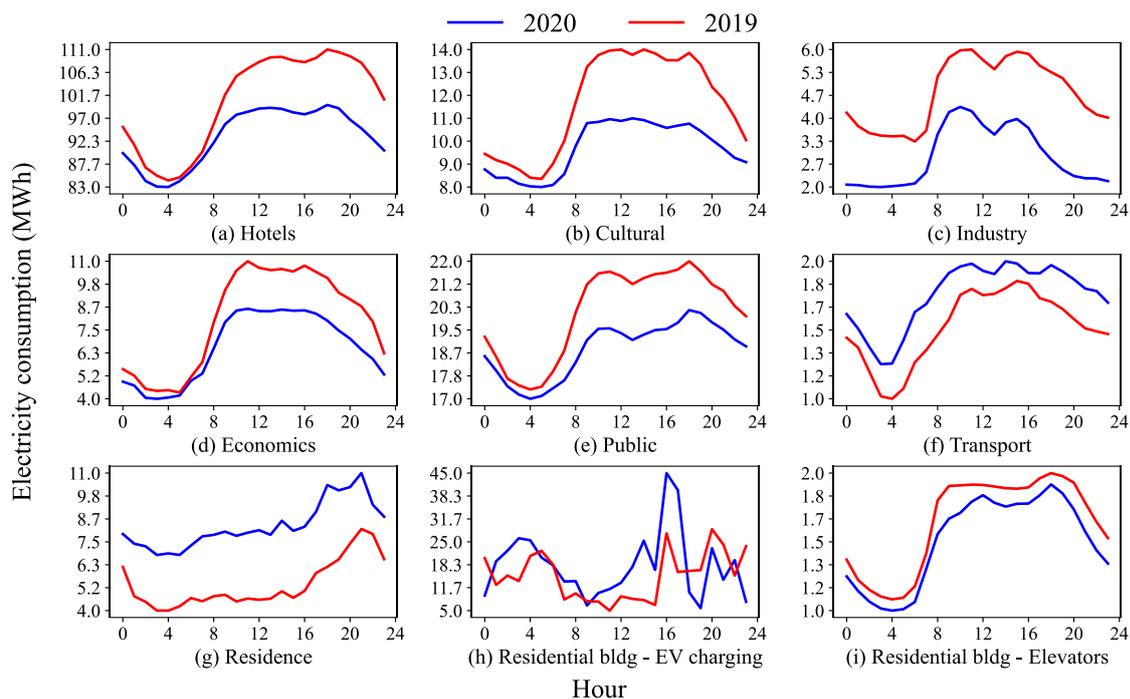

Fig. 10 Hourly average electricity consumption of different sectors, from 27th Jan to 9th Feb in 2019 and 2020

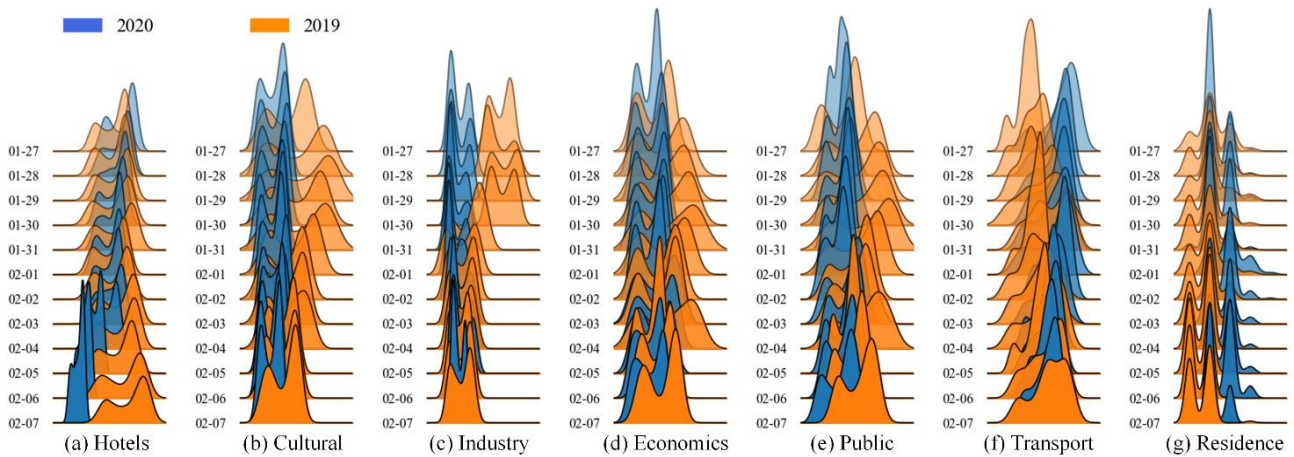

Fig. 11 Daily electricity consumption distribution of different sectors, from 27th Jan to 9th Feb in 2019 and 2020

### 3.1. Energy Consumption Characteristic of Hotels

As it is shown in Fig. 8(a), in 2020, the electricity demand of Hotels (mainly contributed by Casino Hotels) shows a significant decline in the end of January, which is distinctly different from the slow-growing trend observed in 2019. Such decline starts at January 24th, right after the citywide lockdown was announced in Wuhan, and the electricity consumption continues to drop after restriction on entering casinos was implemented by the local government. the influence of the second breakout is also reflected in Fig. 8(a), where the blue line dives in early April. In the end of May 2020, the electricity demand recovered to a relatively high level. The distribution of electricity demand in Fig 9 (a) implies that the expectation of the electricity demand in 2020 is lower than it is in 2019, while the variances are similar. As shown in Fig. 10 (a), after January 27$^{th}$ when the restriction was announced, although the shape of average load curve was similar, the peak load from 10:00 a.m. to 10 p.m. dropped by around 15%, leading to a significant reduction in total electricity consumption. This is also implied by the distribution of daily electricity demand in Fig 11 (a), where the peaks of the ridgelines of 2020 move to the left after restriction announced. Furthermore, the load curve was flattened during the restriction, which is shown by Fig 10 (a) and Fig 11 (a), indicating lower stress during peak hours.

*3.2. Energy Consumption Characteristics of Cultural Events*

Significant reduction in electricity demand from late January is also seen in the cultural events, which consists of education, scientific and recreational activities. In order to prevent people from gathering, the development of remote teaching and online entertainment activities kept the electricity demand at a relatively low level for about three months. After the schools started to reopen from May 4$^{th}$, the electricity demand grows gradually and exceeds the level before COVID-19 pandemic. Fig. 8 (b) shows that the electricity demand in the first few days in May 2020 is remarkably higher than it was in 2019, resulting from the cancellation of the Labour Day holiday and the preliminary preparation for reopening the campus. As shown in Fig. 9 (b), in 2020 the expectation and the variance of electricity both decreased because of the outbreak of the COVID-19 pandemic. After the restriction was announced, the daily peak demand of cultural events reduced by more than 20%, thus flattened the load curve. As indicated by Fig. 8 (b) and Fig. 10 (b), the electricity consumption in late January in 2020 was lower than it was in 2019 before the outbreak of the pandemic, mainly due to the earlier advent of Spring Festival in 2020, thus the impact of the implementation of restriction measures on energy consumption in cultural events is less obvious that it is in hotels. The ridgelines in Fig. 11 (b), which show the variance of the distribution of hourly electricity consumption, narrowed both in 2019 and in 2020, indicating the implication of the Spring Festival. However, the demand level in 2020 was lower than 2019, which reflects the influence of the COVID-19 pandemic.

*3.3. Energy Consumption Characteristics of Industry*

The electricity consumption of the industry is not obviously influenced by the pandemic, since the difference between 2019 and 2020 is not as large as shown in other fields. In Fig. 8(c), the demands show an markable decrease in late January, resulting from the Chinese traditional holiday, the Spring Festival in 4 Feb 2019 and 24 Jan 2020. However, in 2019, a rising trend appeared soon after the Spring Festival, while the electricity consumption remained low in 2020. The electricity demand recovered to the original level from March, but the demand in 2020 was a bit lower than it was in 2019. According to Fig. 9 (c), the maximum demand in 2020 is a little lower than it was in 2019, whereas the minimum demand remains at the same level. However, due to the outbreak of the pandemic, the electricity demand in 2020 concentrate on a lower level than 2019, indicating the decrease in peak demand. As suggested in Fig. 10 (c), a reduction in electricity consumption was observed after the restriction was announced, while the depression of overall demand level in the following two weeks was dominated by the Spring Festival.

### 3.4. Energy Consumption Characteristics of Economic Events

The economic events include the commercial buildings and activities, bank operations and trade. Decrease in electricity demand in 2020 is observed in the two times of breakout of COVID-19, in the end of January and the beginning of April, respectively. The reduction of electricity demand was observed before the announcement of restriction measures, which can be explained by the spontaneous behaviour of the publics to avoid gathering in commercial buildings and banks. The influence of the cancellation of the Labour Day holiday on the energy reviving process in May can also be observed in Fig. 8 (d). Despite this, the overall electricity consumption level in 2020 is similar to what it was in 2019, resulting in the similarity of electricity demand distribution in Fig. 9 (d). However, the impact of the restriction is obvious in the two weeks after its implementation. The demand in the peak hours reduced by around 30%, while the off-peak demand remained at the same level, as shown in Fig. 10 (d) and Fig. 11 (d).

### 3.5. Energy Consumption Characteristics of Administration

The electricity consumption of the public administration shows an obvious decrease in late January. The servants in the government followed strict work-from-home policy when the pandemic broke out, and different from the employees who take turns to go to their office in order to minimize the number of people gathering in the working space. Hence the difference between the electricity demands of 2019 and 2020 remains large from January to May. Gradually recovered from the beginning of May 2020, the electricity demand exceeded the level in 2019 in the middle of May and remained in the same level in the following days, resulting from the cancellation of the Labour Day holiday. The continuous depression of electricity consumption is also reflected by Fig. 9 (e), which shows that the expectation of the electricity demand in 2020 is lower than in 2019. The demand level decreased after the implementation of restriction and work-from-home policy, while the load curves of two years are in the same shape, indicating that although there are less staff working in the office, the basic load pattern was not changed by those policies, as shown in Fig. 10 (e) and Fig. 11 (e).

*3.6. Energy Consumption Characteristics of Transportation*

In the beginning of 2020, the electricity consumed by transportation was remarkably higher than the same period in 2019. Effected by the COVID-19 pandemic breakouts, substantial transportations including flights, ships and coach were stopped, resulting in the electricity consumption-drop in the end of January and in early April. However, since the higher level of the electricity consumption at the beginning of 2020, the average electricity demand in 2020 slightly exceeds the level in 2019, as shown in Fig. 9 (f). During the pandemic, people prefer to stay at home, which is consistent with the observation in the economic sector, resulting in the decrease of electricity consumption after the implementation of restriction, as shown in Fig. 10 (f) and Fig. 11 (f). This can also explain why the electricity demand in 2020 does not show the uptrend as it did in 2019.

*3.7. Energy Consumption Characteristic of General Residence*

Different from other fields, a significant increase of electricity consumption can be observed in the residential buildings after the first pandemic breakout in late January. As shown in Fig. 8 (g)~(i), electricity consumption reductions are found in stairs/elevators and electrical vehicle charging of residential buildings, which are also evidences of the stay-in-home tendency after the restriction measures were announced. This is also supported by Fig. 9 (g), Fig. 10 (g)~(i) and Fig. 11 (g), which indicate that in a short period after the implementation of restriction, people tended to be staying at home, including working-from-home and going off-duty earlier, resulting in the substantial increase in the electricity demand all day long. However, since such tendency did not last long, the distribution of electricity demand from January to May in 2020 is similar to the one in 2019. Such situation only last for a relatively short period of time, since most of the companies did not apply work-from-home policy as strict as the public administration did. Another increase in household electricity demand was observed when the pandemic broke out again in early April. Similar trends also appear in the stairs/elevators and electrical vehicle charging of residential buildings.

*3.8. Summary of Energy Consumption Characteristic of Different Sectors*

In general, after the COVID-19 pandemic broke out, the electricity consumption changes in different ways in different sectors, indicating the varying response patterns of these sectors. Some of the sectors, for example, the hotels and the economics, the reaction to the pandemic might appear before local implementation of the restriction measures. For cities like Macau, which highly relies on the tourism industry, the outbreak of pandemic in neighbouring regions will have an impact on the industries related to tourism (e.g. hotels and commercial buildings), regardless of the local policy. This is a spontaneous reaction by the publics to avoid getting infected. The announcement of restriction by the local government will lead to a further depression on these industries, and the situation only improves when the local restriction is relaxed and the pandemic in the adjacent regions is under control.

Additionally, since the work-from-home policy is always implemented to contain the spread of the virus, the electricity demand in residential area is likely to increase when the pandemic breaks out. At the same time, the electricity demand in office and transport will decrease. The periods of low demand vary from different sectors. Although work-from-home policy was announced after the outbreak, commercial loads and industrial loads tend to recover as soon as the pandemic is well controlled locally, while the depression of electricity demand lasts longer in public administration. As observed in Macau, the servants conducted remote-working for a significantly longer period than the people who work for enterprises. Similar trend could also be found in education and other cultural events. Students are required to take remote courses, until the pandemic is controlled both locally and in neighbouring regions. Therefore, the electricity demand from educational institutes will be kept in a low level for longer time than other sectors, and is the most likely one to decrease with a second pandemic outbreak.

4. **The Energy Consumption Characteristics and Energy Structure of the Commercial Tourism Cities Like Macao**

   *4.1. Analysis of Energy Consumption Characteristics*

   Compare the electricity consumption data of Italy, the United States, Japan and Brazil from January to May 2019 and 2020 [30]. It can be seen from the Fig. 12 that, affected by the COVID-19 pandemic, the implementation of the city closure policy has greatly reduced the activity of many large-scale industries and commerce, tourism and service industries. As a highly developed capitalist country of the four major European economies, Italy's electricity demand from March to May 2020 was significantly reduced. The United States, the world's largest superpower, had a downward trend in power demand from January to April. However, compared with April 2020, the electricity consumption of the two countries in May showed signs of recovery. Due to factors such as land area, population and climate, the electricity demand of the United States is not only higher than that of Italy, but also exceeds that of itself in the same month of last year. This is due to the outbreak of the epidemic with the continuous control of the exhibition and the gradual implementation of the plan to resume work and production in May. Compared with 2019, Brazil, as the largest developing country in South America, is affected by the epidemic situation. In 2020, its power demand decreased month by month from January to May. Such reduction was alleviated in May, which indicates that Brazil's closed epidemic control policy has been eased. However, the data shows that the change of electricity demand in Japan is not obvious, which might be determined by the attitude and measures taken by the Japanese government in the face of the epidemic.

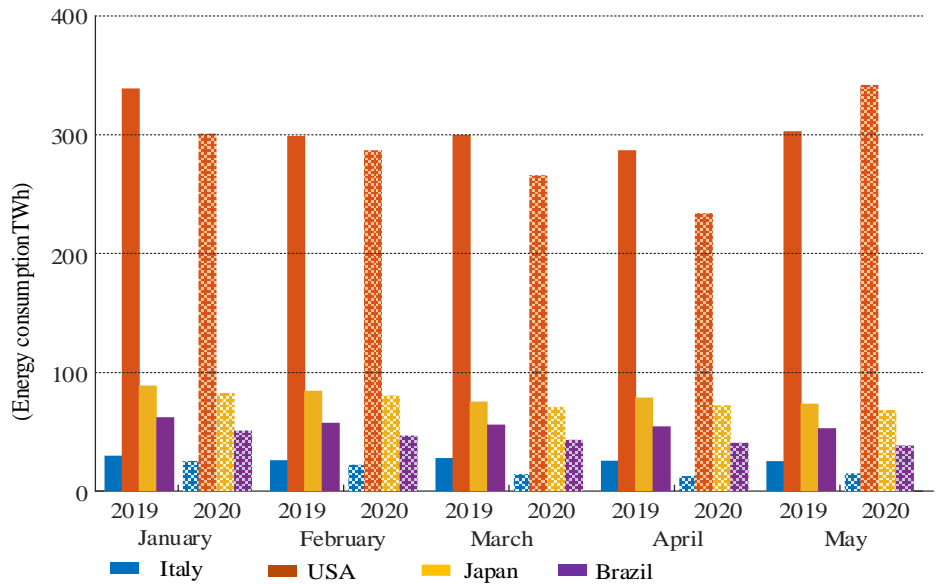

Fig. 12 Energy consumption during epidemic in typical countries

Compared with the above analysis of Macao's electricity consumption data, due to urban closure and other reasons, Macao's electricity consumption trend in the first quarter of 2020 is the same as that of Italy, the United States and Brazil. Under the influence of the COVID-19 pandemic, the overall energy demand showed a downward trend. With the fast control of the epidemic and the gradual implementation of the plan to resume work and production, power demand began to rise in May. As a city where many world-class casinos were located [31], Macao's economy mainly relies on tourism, gambling, light industry and foreign trade [32]. Therefore, Macao's main energy consumption is also centered around tourism and service industries and its commercial power consumption accounts for 64.3% of the total. As one of the most densely populated cities [33], Macao's residential electricity accounts for 22.3%. Compared with the above-mentioned countries led by industry, Macao's tourism service industry also makes the overall energy consumption change relatively gentle under the impact of the pandemic [34]. Due to the large demand of industrial customers, the shutdown of factories during the epidemic will bring greater impact on the energy consumption characteristics. Compared with the large comprehensive cities, Macao's overall energy use is unique in the following aspects:

1) The famous entertainment industry

Just like Macao's famous gambling industry [35-36], typical commercial and tourism cities have their own unique entertainment industries. As an important resource, power supply realizes air conditioning, illumination and various equipment in entertainment venues. Therefore, the entertainment industry has an important impact on the energy and power consumption characteristics of commercial tourism cities. Especially affected by the COVID-19 pandemic, many cities have adopted measures such as city blockades and road closures, which have dealt a huge blow to the entertainment industry. This has also affected the characteristics of the overall electricity consumption of a typical commercial tourism city like Macao, where the overall electricity consumption reduced.

2) High quality hotel service industry

As a commercial tourism city, just as Macao has developed an excellent service industry due to its unique geographical location and gaming culture [37-38], its hotel service industry as a support for the tourism industry is also the key to attract a large number of tourists. The gorgeous hotel architecture and high-quality internal services attract thousands of tourists every year [39]. Due to its unique operating characteristics and the nature of residents, hotel facilities also account for a large part of the overall power consumption of commercial tourism cities. To retard the spread of the COVID-19 pandemic, the hotel service industry has entered the rock bottom, causing the income and energy consumption of such cities to decline [40]. Moreover, the quarantine measures during the epidemic also made some hotels become quarantine areas, offsetting the reduction in electricity consumption.

3) Rapid development of foreign trade

As a member of the World Trade Organization, Macao maintains good trade relations with many countries and regions, mainly including light industrial products and services trade [41]. At the same time, with the vigorous development of tourism, the economic status of service trade in commercial tourism cities is becoming more and more important. As a result of the spread of COVID-19 pandemic, more than 200 countries in the world have been infected. It has a huge impact on global trade and hindered the development of Macao's foreign trade, and thus seriously affected Macao's energy consumption characteristics.

*4.2. Analysis of Power Energy Structure*

According to IEA data, the share of renewable energy in the power structure is still large during COVID-19 pandemic, which does not exclude the regional seasonal climate impact. Due to the low power demand during the epidemic period, considering the low operating cost of renewable energy and the priority of grid access, the power structure in many countries is gradually turning to renewable energy. Fig. 13 shows the energy structure of several typical countries during the epidemic.

In the United States, natural gas is the main source. Due to the implementation of restrictive measures, renewable energy has contributed far more to electricity than coal-fired power plants. In March 2020, despite the reduction of seasonal wind power generation and the easing of the severity of government anti-epidemic measures, natural gas continues to dominate. After the implementation of the blockade measures in India, the gap between coal and renewable energy has been significantly narrowed. The coal remained below 70% of the power structure. In the end of May, power demand gradually recovered and the share of renewable energy continued to rise. Since the end of June, with the increase of temperature and power demand, the proportion of coal in the power structure has increased while wind energy has been declining. Due to beneficial weather conditions and geographical location, EU countries are abundant in renewable resources such as wind power. Their power generation has increased significantly compared with the first quarter of 2019. Affected by the COVID-19 pandemic, the reduction of electricity demand and the increase of renewable energy production further deepened the decline of coal and nuclear energy demand. From June to July, the increase of natural gas demand was second only to renewable energy. In the case of low nuclear power production, natural gas made up for the fluctuation of wind power output every week.

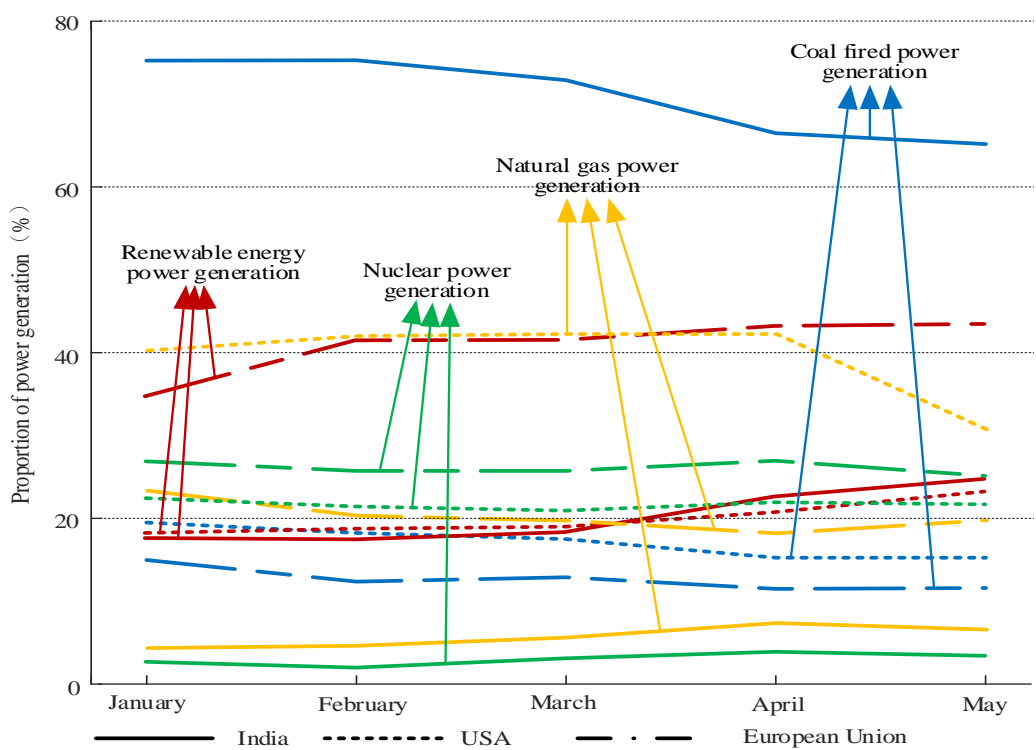

Fig. 13  Energy structure of typical country

Generally speaking, a commercial tourism city like Macau is characterized by dense population and scarce resources. It relies heavily on external resources to meet energy needs. For example, the three major sources of electricity in Macau are local power plants, municipal solid waste incineration power generation, and electricity imported from Mainland China [42-43]. Usually due to the unique geographic location of such cities, renewable resources account for only a small proportion of the energy structure. In addition, the development of the tourism service industry has also brought greater potential for waste incineration power generation in commercial tourism cities like Macao [44]. Generally, the waste incineration power generation capacity of such economic tourism cities is higher than the local natural gas and oil power generation capacity. Therefore, this is also a relatively unique part of the energy structure of this type of cities, and it is a relatively effective method for the treatment and utilization of much garbage in modern cities.

5. **Conclusion**

This paper analyzes the energy characteristics of the urban energy consumption of the commercial tourism city under COVID-19 pandemic based on the energy supply and consumption data of Macao. Compared with 2019, the paper expounds the overall macro trend of energy supply and consumption in Macao. The situation of energy use in various fields in Macao during the COVID-19 pandemic period was analyzed in detail. Compared with the energy consumption and structural characteristics of some typical countries, the energy usage characteristics of commercial tourism cities are analyzed, taking Macao as a representative. The main conclusion of this paper are as follows:

(a) Affected by the COVID-19 pandemic, the energy supply and consumption of commercial tourism cities represented by Macao are showing a downward trend. The energy demand of residential users has been increasing under the influence of the blockade policy. The discharge of waste decreased sharply in a short time, which reduced the environmental pollution.

(b) The energy consumption of hotels, cultural events, shopping malls, entertainment and other service industries and public utilities of a typical commercial tourism city like Macao is obviously affected by the epidemic. The electricity consumption of the industry is not obviously influenced by the pandemic.

(c) Macao is short of resources and its energy supply is mainly rely on Mainland China. During the new outbreak, Macao's natural gas utilization increased. Due to the characteristics of economic development and geographical resources, the current energy structure of most commercial tourism cities in China is still dominated by coal and oil, and is constantly transforming to a renewable energy structure.

**Acknowledgements**

This paper is funded by The Science and Technology Development Fund, Macau SAR (File no. 0137/2019/A3). The authors would like to thank the colleagues from the Companhia de Electricidade de Macau and the Nam Kwong Natural Gas Cooperation for providing anonymous energy consumption data and valuable research advices.